\documentclass[%
 reprint,
 amsmath,amssymb,
 aps,
pra,
]{revtex4-1}

\usepackage{subfigure}
\usepackage{graphicx}
\usepackage{dcolumn}
\usepackage{bm}
\usepackage{amsmath,amsfonts,amssymb}

\begin{document}


\title{Dynamical blockade in a bosonic Josephson junction using optimal coupling}

\author{Dionisis Stefanatos}
\email{dionisis@post.harvard.edu}

\author{Emmanuel Paspalakis}

\affiliation{Materials Science Department, School of Natural Sciences, University of Patras, Patras 26504, Greece}

\date{\today}

\begin{abstract}

In this article we use time-dependent Josephson coupling to enhance unconventional photon blockade in a system of two coupled nonlinear bosonic modes which are initially loaded with weakly populated coherent states, so the evolution is restricted to the manifold of up to two field quanta. Using numerical optimal control, we find the optimal coupling which minimizes the two-photon occupation of one mode, which is actually transferred to the other, while maintains a non-zero one-photon occupation in the same mode. Moreover, we choose the continuous coupling to vanish after the transfer between the modes such that they are decoupled and one of them is left only with some one-photon population which can be observed upon its decay. We numerically find lower values of the second-order correlation function obtained at earlier times than with constant coupling, with larger one-photon populations and for longer time windows, corresponding thus to higher emission efficiency and easier detection. The presented methodology is not restricted to the system under study, but it can also be transferred to other related frameworks, to find the optimal driving fields which can improve the single-photon emission statistics from these systems.

\end{abstract}

\maketitle

\section{Introduction}

\label{sec:intro}

Photon blockade is a nonlinear quantum phenomenon which favors the one-photon state of a quantum mode, while preventing the formation of multi-photon states \cite{Imamoglu97,experiment}. This effect can be exploited for the generation of single photons, a key procedure at the heart of many modern quantum technology applications \cite{Review}, and for implementing quantum simulation and many-body physics with light \cite{Plenio06a,Greentree06a,Angelakis07a,ReviewAngelakis}. Conventional photon blockade is based on the presence of a strong nonlinearity which leads to anharmonic spectra of multi-photon states, allowing thus the selective excitation of the single-photon state \cite{Imamoglu97}. The method fails when the nonlinearity is weaker than the linewidth.

In 2010 Liew and Savona, using a system of coupled quantum modes, showed that unconventional photon blockade can be achieved even for weak nonlinearity \cite{Liew10}. As explained in Ref. \cite{Bamba11a}, this effect is based on the destructive interference between two excitation paths of the two-photon state. An alternative explanation was given in Ref. \cite{Lemonde14}, as the destructive interference between squeezing and displacement of a Gaussian state. Closely related is the understanding of this interference phenomenon in terms of homodyning the signal \cite{Vogel95}, an idea which has been successfully used recently \cite{Muller16,Fischer17}, see also the review \cite{Casalengua}. Several variations have been suggested in order to improve unconventional photon blockade \cite{Ferretti10,Bamba11b,Flayac15,Shen15,Shen15b,Flayac17a,Flayac17b,Shen17,Ghosh18,Shen18}, while other blockade schemes based on similar ideas have been proposed \cite{Gerace14,Kyriienko14,Kryuchkyan16,Cheng17,Sarma17,Sarma18,Li19,Li19b}. Recently, the unconventional photon blockade was implemented experimentally \cite{Vaneph18,Snijders18}. The drawback of this blockade mechanism, also encountered in these experiments, is that the two-time second-order correlation function presents rapid oscillations, thus its detection requires high time resolution. In order to overcome this problem, Ghosh and Liew introduced recently the dynamical blockade scheme \cite{Ghosh19}. According to this method, a combination of continuous and pulsed excitations is applied to a single nonlinear bosonic mode, resulting in a much stronger blockade during longer time windows.

In the present work we use a dynamical (time-dependent) Josephson coupling in order to enhance unconventional photon blockade in the standard framework of two coupled nonlinear bosonic modes \cite{Flayac17a}. We consider that initially the two modes are loaded with weakly populated coherent states, as in Ref. \cite{Flayac17a}, thus the system evolution is approximately restricted to the manifold of up to two field quanta. Using numerical optimal control we find the time-dependent coupling which minimizes the two-photon occupation of one mode (by sending it to the other), while maintains a non-zero one-photon occupation in the same mode. Additionally, the continuous coupling is selected such that it vanishes after the transfer between the modes is completed. The two modes are thus decoupled and one of them is left with some one-photon population, which can be detected upon its decay. We show with specific examples that when using an optimal time-dependent coupling, lower values of the second-order correlation function can be obtained at earlier times than with constant coupling, as in Ref. \cite{Flayac17a}. Moreover, these results are accompanied by larger one-photon populations and are maintained for longer time windows, corresponding thus to higher emission efficiency and easier detection. The suggested methodology is not restricted to the system studied, but can also be exploited in other related contexts, for example the original dynamical blockade framework of a driven dissipative nonlinear bosonic mode \cite{Ghosh19}, to find the optimal driving field which can improve the single-photon emission statistics.

The paper is organized as follows. In the next section we summarize the theoretical framework of the current study, while in Section \ref{sec:results} we present the results. Section \ref{sec:conclusion} concludes this work.

\section{Theory}

\label{sec:theory}

The Hamiltonian describing a bosonic Josephson junction with two quantized modes in the Bose-Hubbard approximation is \cite{Flayac17a,Flayac17b,Stefanatos18a}
\begin{equation}\label{hamiltonianfull}
  H=\sum_{i=1}^2\left[\hbar\omega\hat{a}_i^\dag\hat{a}_i+\hbar U\hat{a}_i^\dag\hat{a}_i^\dag\hat{a}_i\hat{a}_i\right]+\hbar J(t)(\hat{a}_1^\dag\hat{a}_2+\hat{a}_1\hat{a}_2^\dag),
\end{equation}
where $\hat{a}_i, \hat{a}_i^\dag$ are the creation and annihilation operators at site $i$, $\omega$ is the common resonant frequency of both modes and $U$ is the nonlinearity strength. The time-dependent Josephson coupling $J(t)$ will be exploited to control system dynamics. For example, in the context of exciton-polaritons this coupling can be controlled by external electric \cite{Christmann10} or optical \cite{Amo10,Askitopoulos15,Ohadi17,Alyatkin20,Kassenberg20} fields. In the framework of superconducting microwave cavities, the coupling can be varied in ns time-scales \cite{Chen14,Roushan17,Neill18}, which is particularly relevant to the recent experimental implementation \cite{Vaneph18} of unconventional photon blockade. The time evolution of the system is described by the following master equation for the density matrix
\begin{equation}\label{master}
\frac{\partial\rho}{\partial t}=\frac{1}{i\hbar}[H,\rho]+L(\rho),
\end{equation}
where
\begin{equation}\label{Lindblad}
L(\rho)=\sum_{i=1}^2\frac{\kappa}{2}(2\hat{a}_i\rho\hat{a}_i^\dag-\hat{a}_i^\dag\hat{a}_i\rho-\rho\hat{a}_i^\dag\hat{a}_i)
\end{equation}
are Lindblad terms with loss rate $\kappa$.

Initially, the two modes are prepared in a separable product of coherent states
\begin{equation}\label{initial_state}
|\psi(0)\rangle=|\alpha_1\rangle|\alpha_2\rangle,
\end{equation}
where
\begin{equation}\label{coherent_state}
|\alpha_i\rangle=e^{-\frac{|\alpha_i|^2}{2}}\sum_{n=0}^\infty\frac{\alpha_i^n}{\sqrt{n!}}|n\rangle,\quad i=1,2,
\end{equation}
with a small average number of quanta
\begin{equation}\label{weak_pumping}
\alpha^2=|\alpha_1|^2+|\alpha_2|^2\ll 1.
\end{equation}
Initial condition (\ref{initial_state}) is exactly the same as in Ref. \cite{Flayac17a} and we use it here to facilitate comparison. As pointed out there, the initial population imbalance between the modes can be set by driving them with Gaussian laser pulses of varying relative strength. At the end of this section we explain how the methodology presented below can be applied even if the initial state is not the product of coherent states, as long as the low-photon approximation is valid.
Master equation (\ref{master}) is actually derived from a stochastic Shr\"{o}dinger equation with random quantum jumps, which become rare for vanishing occupation numbers of the two modes, as we consider here. In this case, the non-diagonal Lindblad terms $2\hat{a}_i\rho\hat{a}_i^\dag, i=1,2$, can be neglected and the density matrix equation becomes \cite{Eleuch08,Flayac17a,Flayac17b,Carmichael91}
\begin{equation}\label{effective_master}
\frac{\partial\rho}{\partial t}=\frac{1}{i\hbar}[H_{eff}\rho-(H_{eff}\rho)^\dag],
\end{equation}
where the effective non-Hermitian Hamiltonian is
\begin{equation}
H_{eff}=H-i\sum_{i=1}^2\frac{\kappa}{2}\hat{a}_i^\dag\hat{a}_i.
\end{equation}
Then, the density matrix can be factorized as
\begin{equation}
\rho(t)=|\psi(t)\rangle\langle\psi(t)|,
\end{equation}
where state $|\psi(t)\rangle$ satisfies the Shr\"{o}dinger equation
\begin{equation}\label{Efective_Schrodinger}
\frac{\partial}{\partial t}|\psi(t)\rangle=\frac{1}{i\hbar}H_{eff}|\psi(t)\rangle.
\end{equation}

In the weak excitation limit (\ref{weak_pumping}), the system evolution is approximately restricted to the manifold of up to two field quanta and thus the state can be well described by the following truncated wavefunction \cite{Eleuch08,Flayac17a,Flayac17b,Carmichael91}
\begin{eqnarray}\label{state}
|\psi(t)\rangle&=&c_{00}(t)|00\rangle+c_{10}(t)|10\rangle+c_{01}(t)|01\rangle\nonumber\\
               &&+c_{11}(t)|11\rangle+c_{20}(t)|20\rangle+c_{02}(t)|02\rangle,
\end{eqnarray}
where $|ij\rangle$ is the state with $i$ and $j$ quanta in the two modes, respectively.
From Schr\"{o}dinger equation (\ref{Efective_Schrodinger})
we obtain the following differential equations for amplitudes $c_{ij}(t)$,

\begin{equation}\label{zerophoton}
  ic_{00}=0,
\end{equation}

\begin{equation}\label{onephoton}
i
  \left(\begin{array}{c}
    \dot{c}_{10}\\
    \dot{c}_{01}
  \end{array}\right)
  =
  \left(\begin{array}{cc}
    -i\frac{\kappa}{2} & J \\
    J & -i\frac{\kappa}{2}
  \end{array}\right)
  \left(\begin{array}{c}
    c_{10}\\
    c_{01}
  \end{array}\right),
\end{equation}

\begin{equation}\label{twophoton}
i
  \left(\begin{array}{c}
    \dot{c}_{20}\\
    \dot{c}_{11}\\
    \dot{c}_{02}
  \end{array}\right)
  =
  \left(\begin{array}{ccc}
    2U-i\kappa & \sqrt{2}J & 0 \\
    \sqrt{2}J & -i\kappa & \sqrt{2}J \\
    0 & \sqrt{2}J & 2U-i\kappa
  \end{array}\right)
  \left(\begin{array}{c}
    c_{20}\\
    c_{11}\\
    c_{02}
  \end{array}\right).
\end{equation}
Note that in Eqs. (\ref{onephoton}) and (\ref{twophoton}) we have omitted the diagonal terms proportional to resonant frequency $\omega$, since they simply add phase factors $e^{-i\omega t}$ to $c_{10}, c_{01}$ and $e^{-2i\omega t}$ to $c_{20}, c_{11}, c_{02}$, which are eliminated by the absolute value operation in the calculation of second-order correlation function. From the initial state (\ref{initial_state}) and the expansion (\ref{coherent_state}) we find the following initial values for the probability amplitudes
\begin{eqnarray}\label{initial_conditions}
\label{c00} c_{00}(0) &=& e^{-\frac{\alpha^2}{2}},\nonumber\\
\label{c10} c_{10}(0) &=& e^{-\frac{\alpha^2}{2}}\alpha_1,\nonumber\\
\label{c01} c_{01}(0) &=& e^{-\frac{\alpha^2}{2}}\alpha_2,\nonumber\\
\label{c11} c_{11}(0) &=& e^{-\frac{\alpha^2}{2}}\alpha_1\alpha_2,\nonumber\\
\label{c20} c_{20}(0) &=& e^{-\frac{\alpha^2}{2}}\frac{\alpha_1^2}{\sqrt{2}},\nonumber\\
\label{c02} c_{02}(0) &=& e^{-\frac{\alpha^2}{2}}\frac{\alpha_2^2}{\sqrt{2}}.
\end{eqnarray}

\begin{figure}[t]
 \centering
		\begin{tabular}{c}
       \subfigure[$\ $]{
	            \label{fig:concept1}
	            \includegraphics[width=0.9\linewidth]{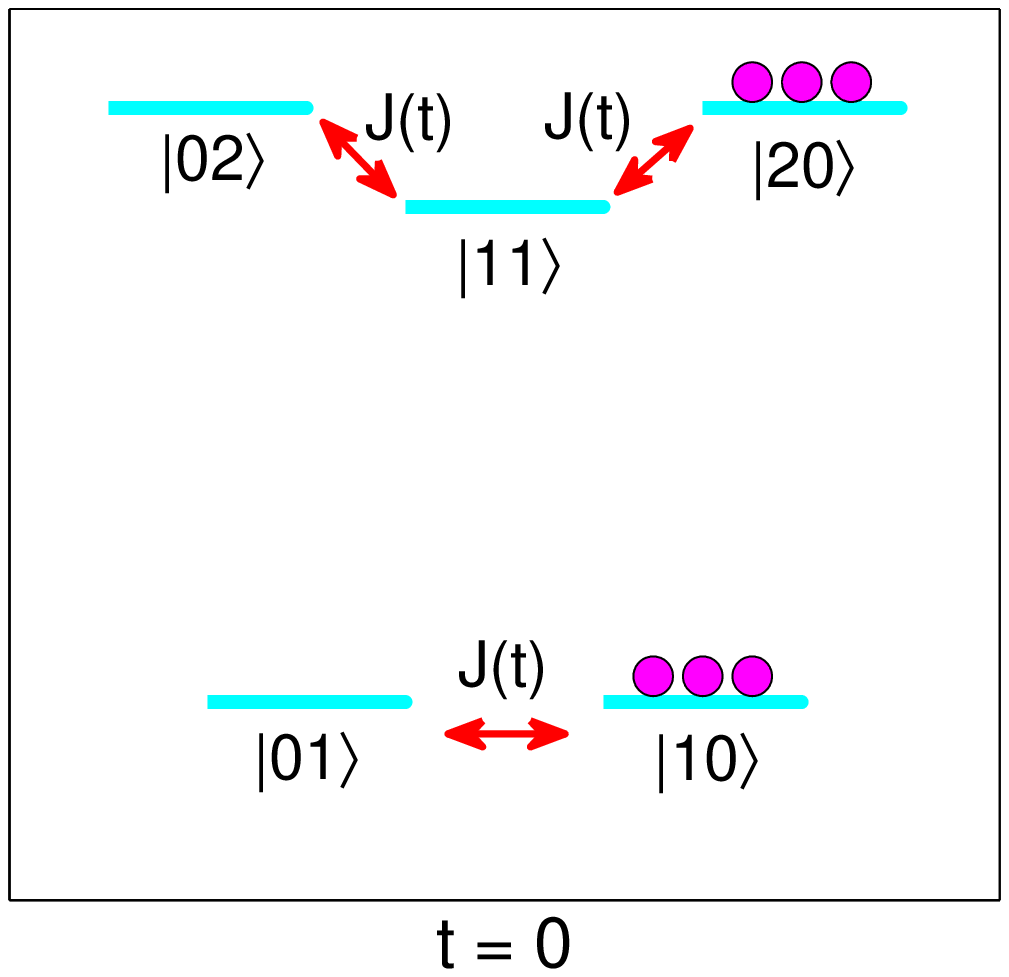}} \\
       \subfigure[$\ $]{
	            \label{fig:concept2}
	            \includegraphics[width=0.9\linewidth]{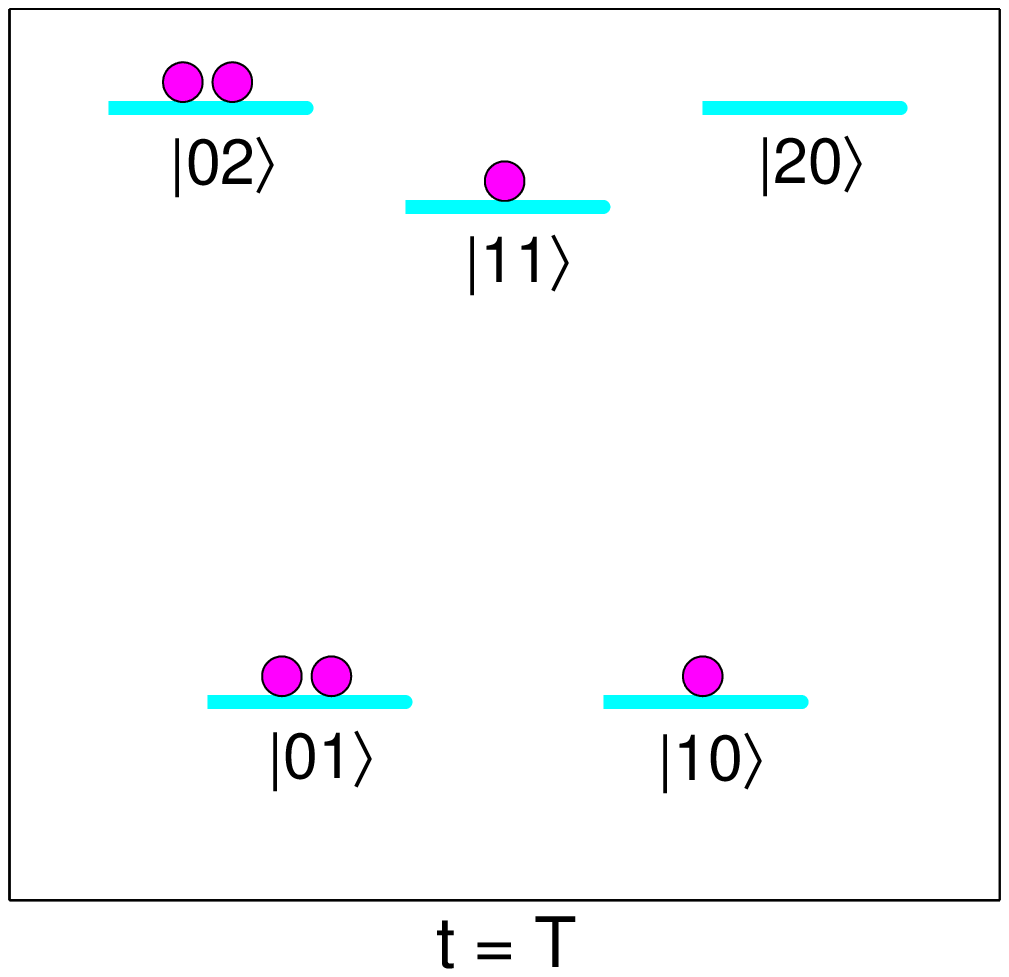}}
		\end{tabular}
\caption{(a) Initially ($t=0$), both states $|10\rangle$ and $|20\rangle$ are populated. (b) Using an optimal time-dependent Josephson coupling $J(t)$, at time $t=T$ most population of state $|20\rangle$ is transferred to the other states of the two-photon manifold, while some population remains in state $|10\rangle$. For $t\geq T$ we set $J(t)=0$, thus the one-photon population is trapped in mode 1 until its decay.}
\label{fig:concept}
\end{figure}

In the recent work \cite{Flayac17a}, Flayac and Savona used a constant Josephson coupling to obtain for one of the modes, let's say mode 1, non-classical values of the equal-time second-order correlation function, $g_1^{(2)}(t,t)<1$, where
\begin{equation}
\label{g2}
g_1^{(2)}(t,t)=\frac{\langle a_1^\dag a_1^\dag a_1 a_1\rangle}{\langle a_1^\dag a_1\rangle^2}=2\frac{|c_{20}|^2}{N_1^2}
\end{equation}
and
\begin{equation}
\label{N1}
N_1(t)=\langle a_1^\dag a_1\rangle=|c_{10}|^2+|c_{11}|^2+2|c_{20}|^2.
\end{equation}
In the present work we try to achieve $g_1^{(2)}(t,t)<1$ using time-dependent coupling $J(t)$. Our idea is to use numerical optimal control to find the optimal $J(t)$ which minimizes the two-photon population in mode 1 at a specific time $t=T$, i.e. $|c_{20}(T)|^2$, starting from initial conditions (\ref{initial_conditions}). If the duration $T$ is large enough then $|c_{20}(T)|^2$ attains a very small value, and only one-photon population is left in mode 1. We also impose the boundary conditions
\begin{equation}
\label{J_BC}
J(0)=0,\quad J(T)=0,
\end{equation}
so initially and finally the two modes are decoupled. For $t>T$ we set $J(t)=0$, thus mode 1 maintains its one-photon population until it is lost due to dissipation $\kappa$. The whole concept is illustrated schematically in Fig. \ref{fig:concept}, for the case where only mode 1 is initially populated. We also constrain $J(t)$ between zero and a maximum value $J_{max}$,
\begin{equation}
\label{constraint}
0\leq J(t)\leq J_{max},
\end{equation}
in order to make a fair comparison with the constant control case where $J(t)=J_{max}$ and also to facilitate the convergence of the numerical solver towards the optimal solution.
Note that a direct minimization of $g_1^{(2)}(t,t)$ presents difficulties since the denominator $N_1(t)$ can become very small at certain times.

For the solution of the optimal control problem, we focus on the two-photon subsystem (\ref{twophoton}). We use the optimal control solver BOCOP \cite{bocop}, thus it is necessary to use real variables instead of complex amplitudes. Let
\begin{equation}
A_2=\sqrt{|c_{20}(0)|^2+|c_{11}(0)|^2+|c_{02}(0)|^2}=e^{-\frac{\alpha^2}{2}}\frac{\alpha^2}{\sqrt{2}},
\end{equation}
thus $A_2^2$ is the initial two-photon occupancy, which is a conserved quantity in the absence of loss ($\kappa=0$). If we define the real variables $x_k, k=1,2,\dots,6$, through the normalized complex amplitudes
\begin{equation}
\frac{c_{20}}{A_2}=x_1+ix_2,\quad \frac{c_{11}}{A_2}=x_3+ix_4,\quad \frac{c_{02}}{A_2}=x_5+ix_6,
\end{equation}
then they satisfy the following system of differential equations
\begin{eqnarray}\label{control_system}
\dot{x}_1 &=& 2Ux_2+\sqrt{2}J(t)x_4,\nonumber\\
\dot{x}_2 &=& -2Ux_1-\sqrt{2}J(t)x_3,\nonumber\\
\dot{x}_3 &=& \sqrt{2}J(t)(x_2+x_6),\nonumber\\
\dot{x}_4 &=& -\sqrt{2}J(t)(x_1+x_5),\nonumber\\
\dot{x}_5 &=& \sqrt{2}J(t)x_4+2Ux_6,\nonumber\\
\dot{x}_6 &=& -\sqrt{2}J(t)x_3-2Ux_5.
\end{eqnarray}
The corresponding initial conditions can be found from Eq. (\ref{initial_conditions}) and they are
\begin{eqnarray}
\label{x_initial_conditions}
x_1(0)&=&\frac{1+z_0}{2},\quad x_2(0)=0,\nonumber\\
x_3(0)&=&\sqrt{\frac{1-z_0^2}{2}},\quad x_4(0)=0,\nonumber\\
x_5(0)&=&\frac{1-z_0}{2},\quad x_6(0)=0,
\end{eqnarray}
where note that we have expressed them using the initial population imbalance between the modes,
\begin{equation}
z_0=\frac{\alpha_1^2-\alpha_2^2}{\alpha_1^2+\alpha_2^2}
\end{equation}
for real $\alpha_1,\alpha_2$.

In order to find a time-dependent coupling $J(t)$ satisfying the boundary conditions (\ref{J_BC}), we consider $J$ as an extra state variable, on which we impose these conditions, while we place the control in its derivative. If we additionally exploit a BOCOP feature which allows to express the control as a harmonic series of time, then the corresponding equation in normalized time $\tau=\kappa t$ is
\begin{equation}
\label{J}
\frac{d}{d\tau}\left(\frac{J}{\kappa}\right)=a_0+\sum_{k=1}^p(a_{2k-1}\cos{k\tau}+a_{2k}\sin{k\tau}),
\end{equation}
where $p$ is the number of harmonics used. We use BOCOP solver to find the coefficients $a_k$ which minimize $|c_{20}(T)|^2/A_2^2=x_1^2(T)+x_2^2(T)$ for specific duration $t=T$, while satisfy the boundary conditions (\ref{J_BC}) and the constraint (\ref{constraint}). Having found $a_k$ we can integrate Eq. (\ref{J}) and obtain the coupling $J(t)$. In the next section we consider two examples, for strong and weak nonlinearity $U$, as in Ref. \cite{Flayac17a}. Note that in the formulation of the optimal control problem we have used without loss of generality real coherent field amplitudes $\alpha_1,\alpha_2$. The same methodology can be applied for complex $\alpha_1,\alpha_2$, and even if the initial state is not a product of coherent states, as long as the low-photon approximation is valid. In these cases one has simply to perform the optimization of the coupling $J(t)$ using the appropriate set of initial conditions (\ref{x_initial_conditions}).

\section{Results}

\label{sec:results}

\begin{figure}[t]
 \centering
		\begin{tabular}{c}
       \subfigure[$\ $]{
	            \label{fig:Tp1}
	            \includegraphics[width=0.9\linewidth]{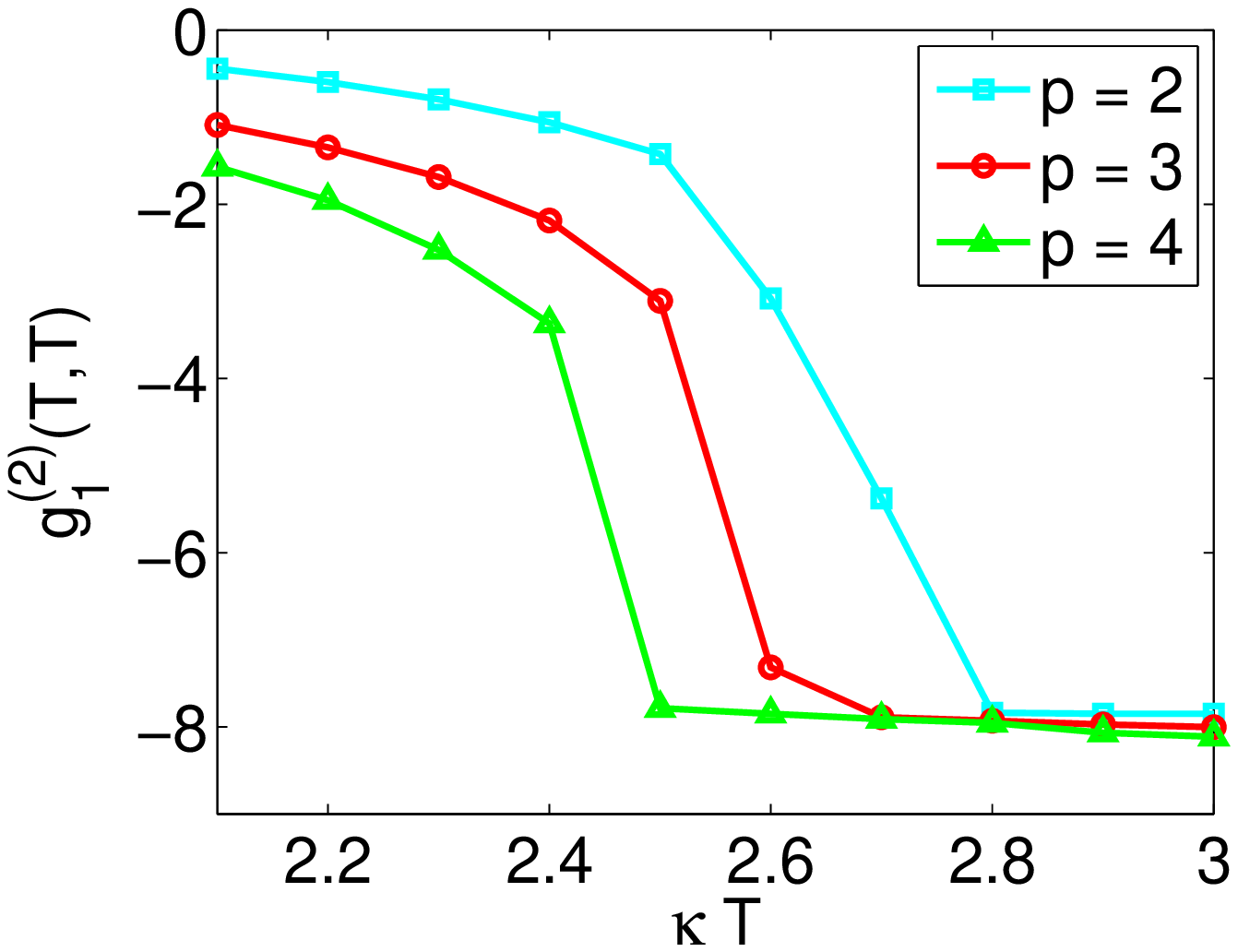}} \\
       \subfigure[$\ $]{
	            \label{fig:Tp2}
	            \includegraphics[width=0.9\linewidth]{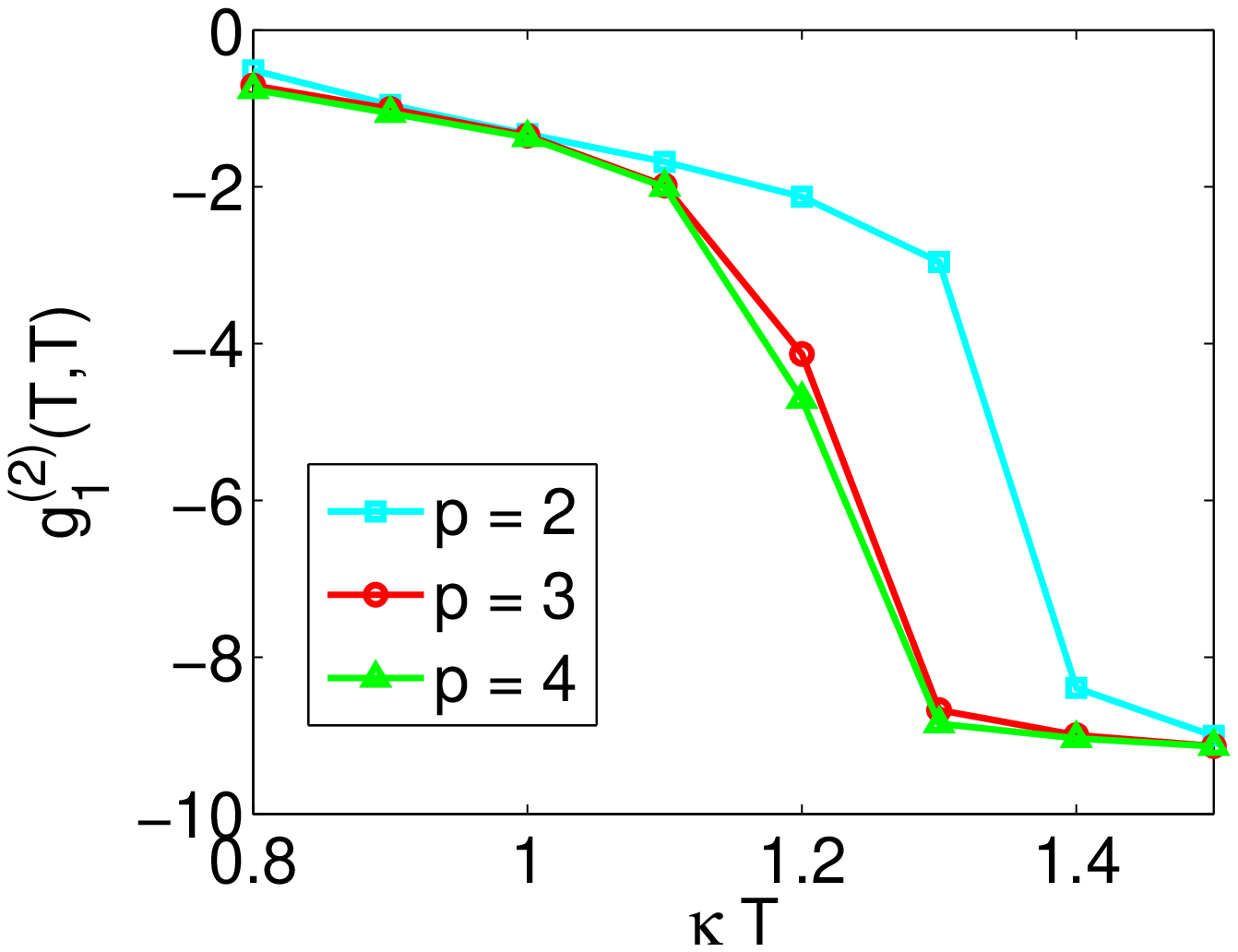}}
		\end{tabular}
\caption{Equal-time second-order correlation function (in logarithmic scale) of the first mode for optimal coupling with duration $T$, when two (cyan squares), three (red circles), and four (green triangles) harmonics are used in the control (\ref{J}). (a) Example with strong nonlinearity $U=\kappa$. (b) Example with weak nonlinearity $U=2\pi\times 10^{-2}\kappa$.}
\label{fig:Tp}
\end{figure}

We study first the case with strong nonlinearity $U=\kappa$. For the other parameters we use the values $J_{max}=5\kappa$, thus $J_{max}=5U$, $\alpha_1=0.1$ and $z_0=1$, so only the first mode is initially populated, as in Ref. \cite{Flayac17a}. In order to find the duration $T$ during which a nonzero $J(t)$ is applied, and the number $p$ of harmonics needed in the control (\ref{J}), we solve numerically the optimal control problem for various values of $T$ and $p$. In Fig. \ref{fig:Tp1} we plot the resultant equal-time second-order correlation function in logarithmic scale, for $2.1\kappa^{-1}\leq T\leq 3\kappa^{-1}$ with step $\delta T=0.1\kappa^{-1}$, and three values of $p$, $p=2$ (cyan squares), $p=3$ (red circles), and $p=4$ (green triangles). For $T=2.6\kappa^{-1}$ and $p=3$ harmonics we find the small value $g_1^{(2)}(T,T)=4.845\times 10^{-8}$, while observe that using more harmonics with this duration increases the complexity of the control without improving much the performance. We thus choose to proceed with these parameter values.

\begin{table}[t!]
\caption{\label{tab:coefficients} Optimal coefficients for the trigonometric series (\ref{J}) with $p=3$. The first column corresponds to the example with strong nonlinearity and the second column to that with weak nonlinearity.}
\begin{ruledtabular}
\begin{tabular}{ccc}
\textrm{Coefficients}&
\textrm{Example 1}&
\textrm{Example 2}\\
\colrule
$a_0$ & 258.3070 & 136.4215\\
$a_1$ & 15.5649 & 5561.9086\\
$a_2$ & -432.1063 & -8295.7429\\
$a_3$ & -236.0900 & -7716.0838\\
$a_4$ & -5.0417 & 2879.9583\\
$a_5$ & 5.3701 & 2081.7869\\
$a_6$ & 57.1314 & 558.8372
\end{tabular}
\end{ruledtabular}
\end{table}

\begin{figure}[t!]
 \centering
		\begin{tabular}{c}
       \subfigure[$\ $]{
	            \label{fig:con1}
	            \includegraphics[width=0.9\linewidth]{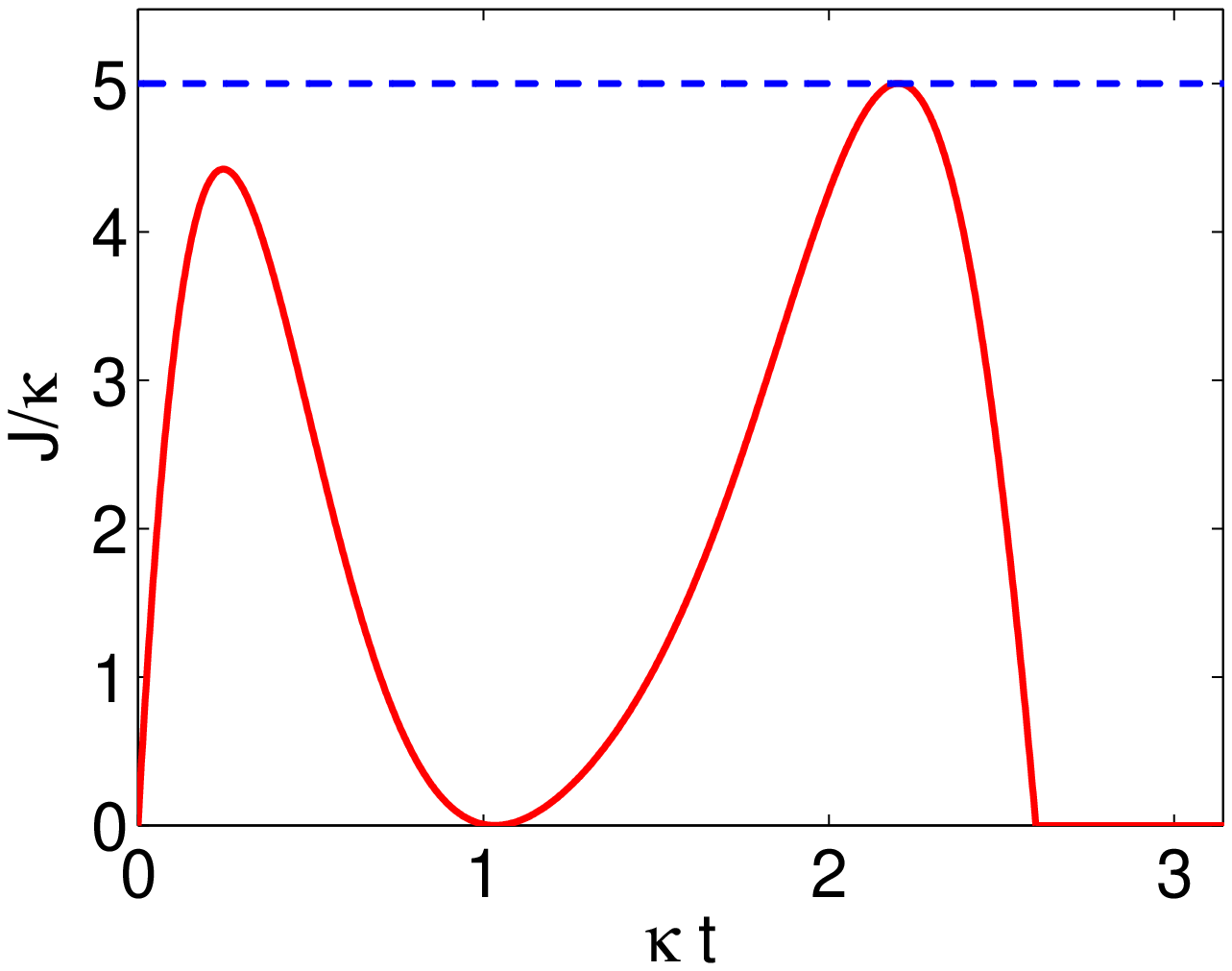}} \\
       \subfigure[$\ $]{
	            \label{fig:g1}
	            \includegraphics[width=0.9\linewidth]{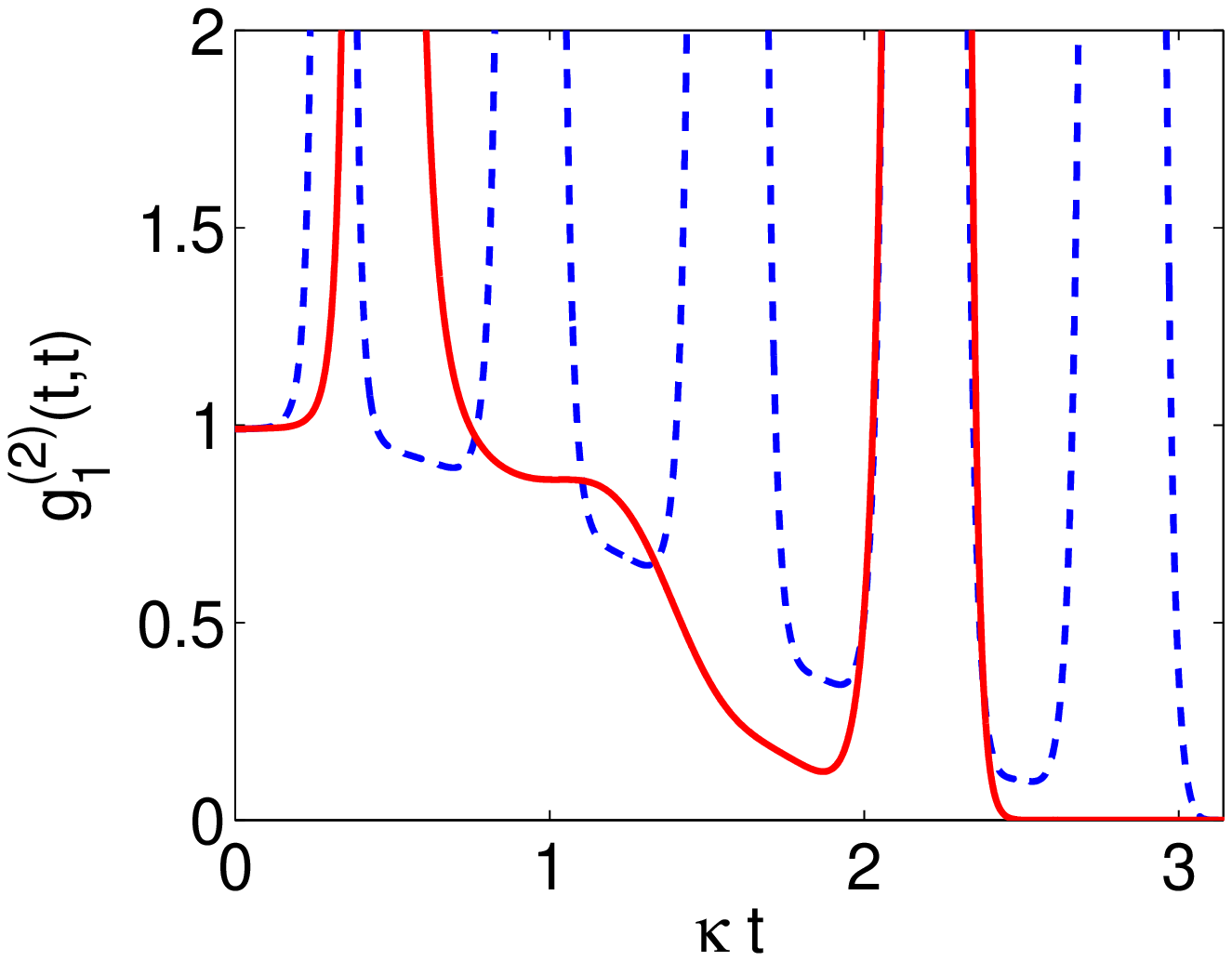}} \\
       \subfigure[$\ $]{
	            \label{fig:pop1}
	            \includegraphics[width=0.9\linewidth]{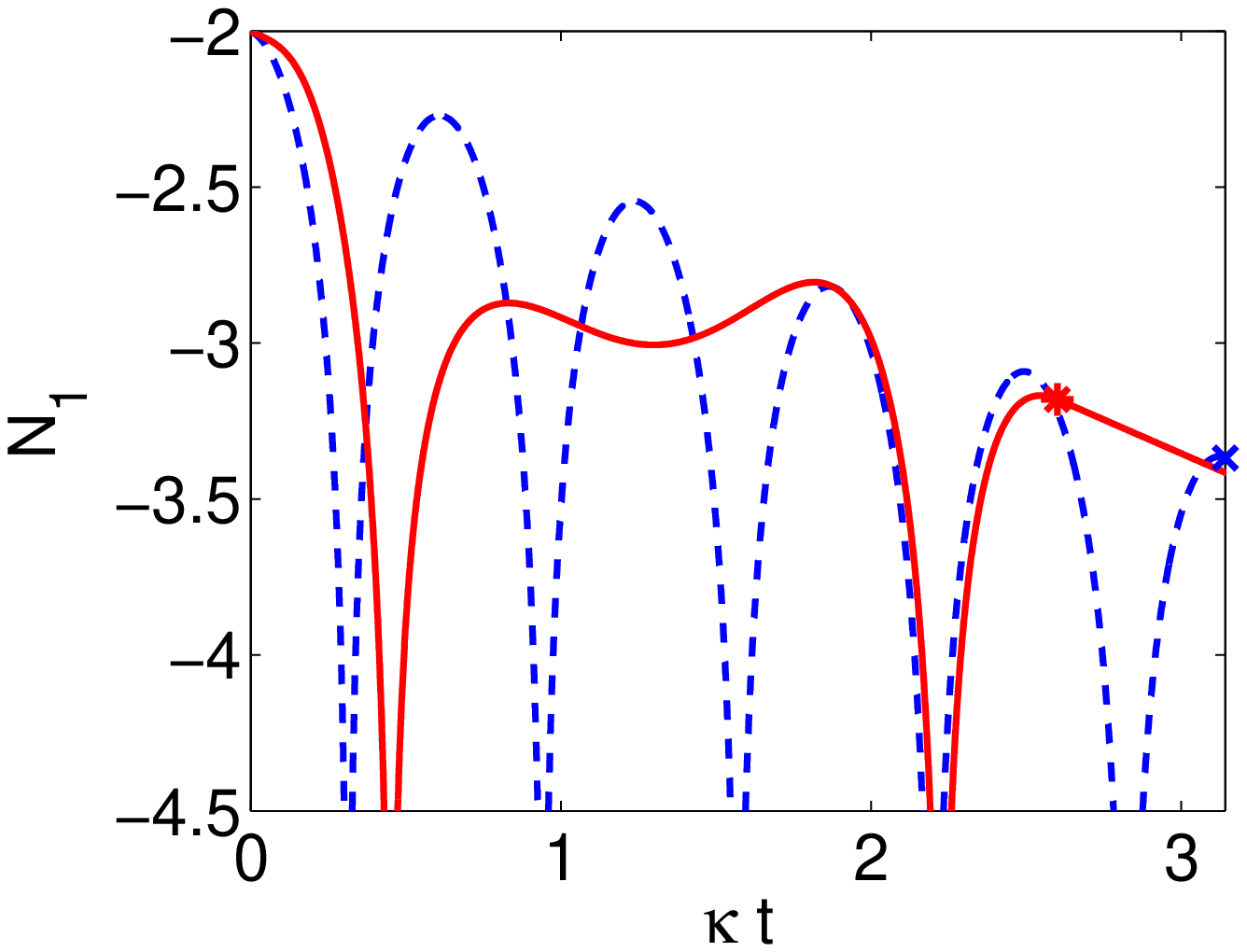}}
		\end{tabular}
\caption{Example for strong nonlinearity $U=\kappa$. Blue dashed line corresponds to constant Josephson coupling $J=5\kappa$ as in Ref. \cite{Flayac17a}, while red solid line to optimal time-dependent $0\leq J(t)\leq 5\kappa$, which is selected to minimize the two-photon occupation $|c_{20}(T)|^2$ of the first mode at $T=2.6/\kappa$ while it vanishes for $t\geq T$. (a) Josephson coupling. (b) Equal-time second-order correlation function for the first mode. (c) Population of the first mode.}
\label{fig:example1}
\end{figure}

For this case, the coefficients $a_k$ of the control (\ref{J}) are displayed in the first column of Table \ref{tab:coefficients}. In Fig. \ref{fig:con1} we plot with red solid line the corresponding Josephson coupling $J(t)$. Note that we have extended time in the interval $T\leq t \leq \pi\kappa^{-1}$, where $J(t)=0$. We also show with blue-dashed line the constant control $J_{max}=5U$ used in Ref. \cite{Flayac17a}. In Fig. \ref{fig:g1} we display the equal-time second-order correlation function for the two cases. With the optimal time-dependent coupling the value $4.845\times 10^{-8}$ is obtained at $t=T=2.6\kappa^{-1}$ and approximately maintained thereafter, when the modes are decoupled. With constant coupling the (larger) minimal value $9.878\times 10^{-5}$ is obtained at the later time $t=\pi\kappa^{-1}$. In Fig. \ref{fig:pop1} we plot in logarithmic scale the population $N_1(t)$ of the first mode for both protocols. Observe that for the time-dependent protocol and most of the interval $T\leq t \leq \pi\kappa^{-1}$, where the corresponding correlation function is minimized, the population is larger than that of the constant protocol at $t=\pi\kappa^{-1}$ (compare the red star marker and the subsequent red straight line with the blue cross marker in Fig. \ref{fig:pop1}). Also, observe from Fig. \ref{fig:g1} that the correlation function of the time-dependent protocol attains non-classical values (lower than one) in a window of approximate width $\kappa^{-1}$, roughly in the interval $\kappa^{-1}\leq t\leq 2\kappa^{-1}$. From Fig. \ref{fig:pop1} we see that the population of mode 1 during this interval is about $10^{-3}$.

\begin{figure}[t!]
 \centering
		\begin{tabular}{c}
       \subfigure[$\ $]{
	            \label{fig:con2}
	            \includegraphics[width=.9\linewidth]{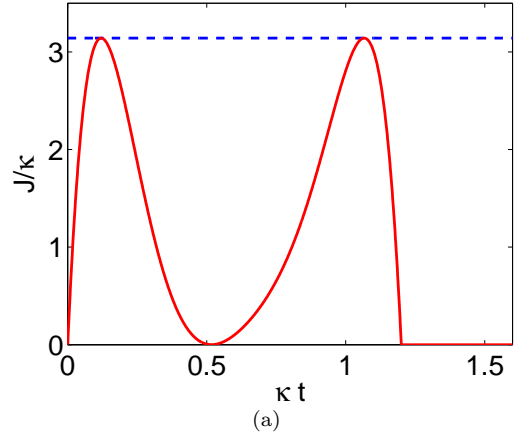}} \\
       \subfigure[$\ $]{
	            \label{fig:g2}
	            \includegraphics[width=.9\linewidth]{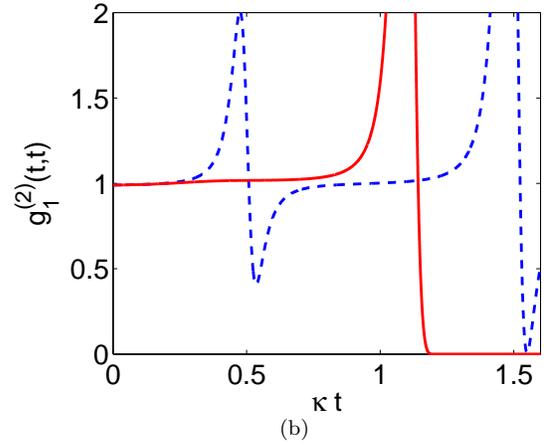}} \\
       \subfigure[$\ $]{
	            \label{fig:pop2}
	            \includegraphics[width=.9\linewidth]{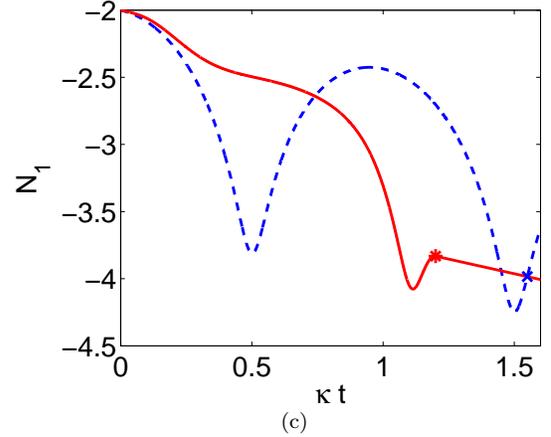}}
		\end{tabular}
\caption{Example for weak nonlinearity $U=2\pi\times 10^{-2}\kappa$. Blue dashed line corresponds to constant Josephson coupling $J=\pi\kappa$ as in Ref. \cite{Flayac17a}, while red solid line to optimal time-dependent $0\leq J(t)\leq \pi\kappa$, which is selected to minimize the two-photon occupation $|c_{20}(T)|^2$ of the first mode at $T=1.2/\kappa$ while it vanishes for $t\geq T$. (a) Josephson coupling. (b) Equal time second-order correlation function for the first mode. (c) Population of the first mode.}
\label{fig:example2}
\end{figure}

The second example that we consider corresponds to a weak nonlinearity $U=2\pi\times 10^{-2}\kappa$, $J_{max}=\pi\kappa$, $\alpha_1=0.1$, and initial population imbalance $z_0=0.95$, as in Ref. \cite{Flayac17a}. As in the previous case, in Fig. \ref{fig:Tp2} we plot the equal-time second-order correlation function in logarithmic scale for various durations in the interval $0.8\kappa^{-1}\leq T\leq 1.5\kappa^{-1}$ with step $\delta T=0.1\kappa^{-1}$, and three numbers of harmonics, $p=2$ (cyan squares), $p=3$ (red circles), and $p=4$ (green triangles). For $T=1.2\kappa^{-1}$ and $p=3$ harmonics we obtain the acceptable value $g_1^{(2)}(T,T)=7.384\times 10^{-5}$, much lower than the minimum value obtained with constant coupling as we shall immediately see, thus we present further results using these values. Obviously, using more harmonics does not improve the performance substantially. The corresponding coefficients $a_k$ of the control (\ref{J}) are displayed in the second column of Table \ref{tab:coefficients}. In Fig. \ref{fig:con2} we plot with red solid line the corresponding Josephson coupling $J(t)$ and with blue-dashed line the constant control $J_{max}=\pi\kappa$ used in Ref. \cite{Flayac17a}, while note that we have extended time in the interval $T\leq t \leq 1.6\kappa^{-1}$, where $J(t)=0$. In Fig. \ref{fig:g2} we display the equal-time second-order correlation function for the two cases. Using the optimal time-dependent coupling, the value $7.384\times 10^{-5}$ is achieved at $t=T=1.2\kappa^{-1}$ and approximately maintained after the modes are decoupled, while using the constant coupling the (much larger) minimum value $8.961\times 10^{-3}$ is achieved at the later time $t\approx 1.55\kappa^{-1}$. In Fig. \ref{fig:pop2} we plot in logarithmic scale the population $N_1(t)$ of the first mode for both protocols. Observe that for the time-dependent protocol and most of the interval $T\leq t \leq 1.6\kappa^{-1}$, where the corresponding correlation function is minimized, the population is larger than that of the constant protocol at $t\approx 1.55\kappa^{-1}$ (compare the red star marker and the subsequent red straight line with the blue cross marker).

It appears from Figs. \ref{fig:g1}, \ref{fig:g2} that, for $t\geq T$, the equal-time second-order correlation function remains constant to its value at $t=T$. We will show that it actually increases in the course of time, although slightly. For $t\geq T$, where the Josephson coupling has been turned off, it is
\begin{eqnarray*}
|c_{20}(t)|^2=&&e^{-2\kappa(t-T)}|c_{20}(T)|^2,\\
N_1(t)=&&e^{-\kappa(t-T)}\times\nonumber\\
&&\left\{|c_{10}(T)|^2+e^{-\kappa(t-T)}\left[|c_{11}(T)|^2+2|c_{20}(T)|^2\right]\right\}\nonumber\\
\leq && e^{-\kappa(t-T)}N_1(T).
\end{eqnarray*}
From the above relations we get
\begin{equation}
g_1^{(2)}(t,t)=2\frac{|c_{20}(t)|^2}{N_1^2(t)}\geq 2\frac{|c_{20}(T)|^2}{N_1^2(T)}=g_1^{(2)}(T,T).
\end{equation}
For example, in Fig. \ref{fig:g2} and for $t_1=1.6/\kappa$, it is $g_1^{(2)}(t_1,t_1)=7.399\times 10^{-5}$, slightly larger than $g_1^{(2)}(T,T)=7.384\times 10^{-5}$. Although
$g_1^{(2)}(t,t)$ is an increasing function of $t\geq T$, observe that for large $\kappa t$ it is $N_1(t)\rightarrow e^{-\kappa(t-T)}|c_{10}(T)|^2\approx e^{-\kappa(t-T)}N_1(T)$, thus the correlation tends to a limit which is larger but close to $g_1^{(2)}(T,T)$.

\begin{figure}[t!]
\centering
\includegraphics[width=1\linewidth]{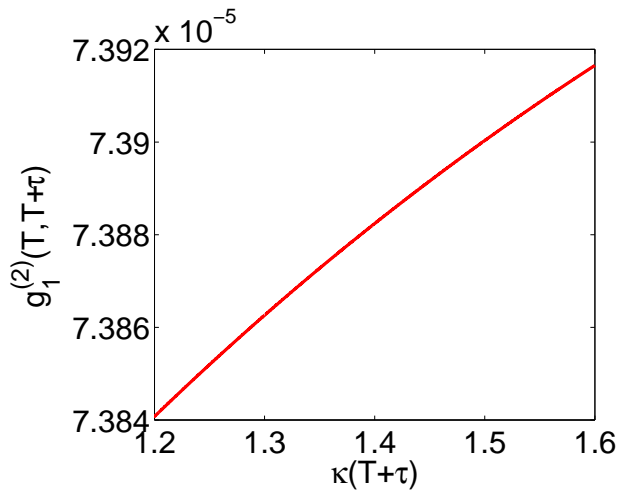}
\caption{Two-time second-order correlation function  for the second example and a photon emission at $t=T=1.2/\kappa$.}
\label{fig:twotime}
\end{figure}

We next find the two-time second-order correlation function $g_1^{(2)}(t,t+\tau)$, for $t\geq T$ and $\tau\geq 0$, following the methodology described in Ref. \cite{Eleuch08}. After the emission of one photon at time $t\geq T$, the wavefunction $|\psi\rangle$ collapses to the reduced state $|\phi_t(0)\rangle=\frac{\hat{a}_1|\psi\rangle}{\langle\psi|\hat{a}_1^\dag\hat{a}_1|\psi\rangle}$. This is a one-quantum state with dynamical behavior similar to $|\psi(t)\rangle$, thus it can be expressed as
\begin{equation*}
|\phi_t(\tau)\rangle=b_{00}(t,0)|00\rangle+b_{10}(t,\tau)|10\rangle+b_{01}(t,\tau)|01\rangle,
\end{equation*}
where $b_{00}$ is constant with respect to $\tau$ while $b_{10}, b_{01}$ obey system (\ref{onephoton}) with initial conditions
\begin{equation*}
b_{10}(t,0)=\frac{\sqrt{2}c_{20}(t)}{\sqrt{N_1(t)}},\quad b_{01}(t,0)=\frac{c_{11}(t)}{\sqrt{N_1(t)}}.
\end{equation*}
Using the reduced state, we express the two-time correlation function as
\begin{eqnarray*}
g_1^{(2)}(t,t+\tau)&=&\frac{\langle\hat{a}_1^\dag(t)\hat{a}_1^\dag(t+\tau)\hat{a}_1(t+\tau)\hat{a}_1(t)\rangle}{\langle\hat{a}_1^\dag(t)\hat{a}_1(t)\rangle\langle\hat{a}_1^\dag(t+\tau)\hat{a}_1(t+\tau)\rangle}\\
&=&\frac{\langle\phi_t(0)|\hat{a}_1^\dag(t+\tau)\hat{a}_1(t+\tau)|\phi_t(0)\rangle}{\langle\hat{a}_1^\dag(t+\tau)\hat{a}_1(t+\tau)\rangle}\\
&=&\frac{\langle\phi_t(\tau)|\hat{a}_1^\dag\hat{a}_1|\phi_t(\tau)\rangle}{N_1(t+\tau)}\\
&=&\frac{|b_{10}(t,\tau)|^2}{N_1(t+\tau)}.
\end{eqnarray*}
For $t\geq T$, after the Josephson coupling has been turned off, we easily find from system (\ref{onephoton}) that
\begin{equation*}
b_{10}(t,\tau)=e^{-\frac{\kappa\tau}{2}}b_{10}(t,0)=\frac{\sqrt{2}e^{-\frac{\kappa\tau}{2}}c_{20}(t)}{\sqrt{N_1(t)}},
\end{equation*}
thus
\begin{equation}
\label{g2ineq}
g_1^{(2)}(t,t+\tau)=\frac{2|c_{20}(t)|^2}{N_1(t)N_1(t+\tau)e^{\kappa\tau}},
\end{equation}
where $N_1(t)$ is given in Eq. (\ref{N1}) and
\begin{equation*}
N_1(t+\tau)=e^{-\kappa\tau}\left\{|c_{10}(t)|^2+e^{-\kappa\tau}\left[|c_{11}(t)|^2+2|c_{20}(t)|^2\right]\right\}
\end{equation*}
for $t\geq T$ and $\tau\geq 0$. Since $N_1(t+\tau)e^{\kappa\tau}\leq N_1(t)$, from Eq. (\ref{g2ineq}) we obtain
\begin{equation}
g_1^{(2)}(t,t+\tau)\geq\frac{2|c_{20}(t)|^2}{N_1^2(t)}=g_1^{(2)}(t,t).
\end{equation}
In Fig. \ref{fig:twotime} we plot the correlation function $g_1^{(2)}(T,T+\tau)$ for the second example considered previously. Observe that, although it increases from its value at $\tau=0$, it remains very small. When $t\geq T$, even for large $\kappa\tau$ it is $N_1(t+\tau)e^{\kappa\tau}\rightarrow |c_{10}(t)|^2\approx N_1(t)$, thus the two-time correlation given in Eq. (27) tends to a limit which is larger but close to $g_1^{(2)}(t,t)$, and thus close to zero. We conclude that if a photon is emitted at $t\geq T$ then, with very high probability, it is the only one. This is intuitively expected since for $t\geq T$ the two-photon population has been transferred to mode 2 and the modes are decoupled. As in the case of dynamical blockade \cite{Ghosh19}, in order to detect single photons a shutter can be placed in the emission and be opened after $t=T$.

\begin{figure}[t!]
\centering
\includegraphics[width=1\linewidth]{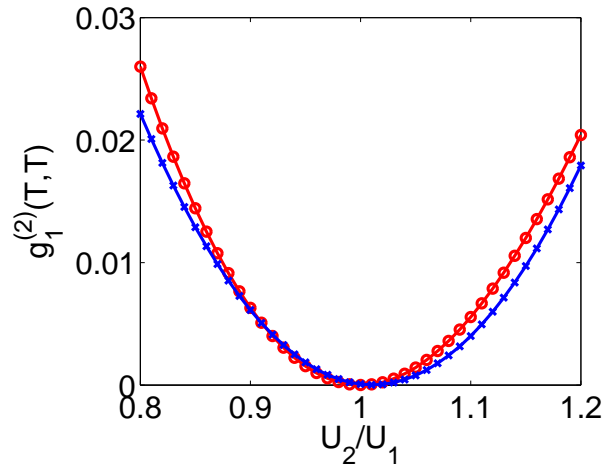}
\caption{Robustness of the equal-time second-order correlation function at $t=T$ when there is a mismatch between the nonlinearities of the modes $U_1$ and $U_2$. Red circles correspond to the first example with strong nonlinearity and blue crosses to the second example with weak nonlinearity.}
\label{fig:robustness}
\end{figure}

In order to test the robustness of the proposed method, in Fig. \ref{fig:robustness} we plot the equal-time second-order correlation function at $t=T$ when there is a mismatch between the nonlinearities of the modes $U_1$ and $U_2$ in the range of $\pm 20\%$. Red circles correspond to the example with strong nonlinearity and blue crosses to that with weak. Observe that for both cases it is $g_1^{(2)}(T,T)\ll 1$. The case with weak nonlinearity appears to be more robust in general, since the deviation from the ideal evolution due to mismatch $U_1\neq U_2$, which is accumulated over time, is smaller. Finally note that for both examples the behavior is better for $U_2>U_1$ than for $U_2<U_1$. Since $U_1$ is held fixed to its unperturbed value, a larger nonlinearity $U_2$ is expected to give better results. Aside to the considered nonlinearity mismatch, other possible experimental limitations include thermal noise, detuning between the modes, and pure dephasing. The effect of these mechanisms on photon statistics for a bosonic Josephson junction with constant coupling has been studied in Ref. \cite{Flayac17a}. The conclusion is that, as long as the strength of these mechanisms is kept below certain levels, the phenomenon of photon antibunching can still be observed. We expect that the same applies in our model, where time-dependent coupling is used.

\section{Conclusion}

\label{sec:conclusion}

In this article we considered the standard framework for unconventional photon blockade with two coupled nonlinear bosonic modes and used an optimized time-dependent coupling to improve single-photon emission statistics from one of the modes. This approach led to lower values of the second-order correlation function at earlier times, with larger one-photon populations and for longer time windows than the case with constant coupling, corresponding thus to higher emission efficiency and easier detection. The proposed methodology can also be applied to other related physical contexts, for example to optimize the driving field in the case of dynamical photon blockade or a time-dependent coupling in Jaynes-Cummings model.

\begin{acknowledgments}
Co-financed by Greece and the European Union - European Regional Development Fund via the General Secretariat for Research and Technology bilateral Greek-Russian Science and Technology collaboration project on Quantum Technologies (project code name POLISIMULATOR).
\end{acknowledgments}

\end{document}